\documentclass[aps,preprint,prd,epsfig,nofootinbib]{revtex4}
\pdfoutput=1
\usepackage{amssymb}
\usepackage{amsmath}
\usepackage{color}
\usepackage{fancyhdr}
\newcommand{\be}{\begin{equation}}
\newcommand{\ee}{\end{equation}}
\newcommand{\bea}{\begin{eqnarray}}
\newcommand{\eea}{\end{eqnarray}}
\newcommand{\bes}{\begin{subequations}}
\newcommand{\ees}{\end{subequations}}

\newcommand{\bc}{\begin{center}}
\newcommand{\ec}{\end{center}}
\usepackage{amsmath, amssymb} 
\usepackage{slashed}
\usepackage{graphicx}
\usepackage{subcaption}
\begin{document}

\title{On the  Higgs spectra of  the 3-3-1 model}

\author{Jo\~ao Paulo Pinheiro$^{a}$, C. A. de S. Pires$^{b}$}
\affiliation{{$^a$ Departament de Física Quàntica i Astrofísica and Institut de Ciències del Cosmos, Universitat
de Barcelona, Diagonal 647, E-08028 Barcelona, Spain}, \\
{$^b$  Departamento de Física, Universidade Federal da Paraíba, Caixa Postal 5008, 58051-970, Jo\~ao Pessoa, PB, Brazil}, \\
}

\date{\today}

\begin{abstract}
The minimal scalar sector of the 3-3-1 model is composed by the SU(3)$_L$ triplet scalars $\eta$, $\rho$, $\chi$ and its potential allows the trilinear term  $\frac{f}{\sqrt{2}}\chi \eta \rho$. Since $f$ is an energy scale associated to the explicit violation of Peccei-Quinn global symmetry, it is natural to consider in what energy scale such symmetry is broken and its consequences in the spectrum of scalars of the model. Here, we show that $f$ determines the spectrum of scalars of the model.  Hence, we develop the scalar sector considering  $f$ belonging to four energy regimes, namely $f \ll  \langle \eta \rangle_0$, $\langle \rho \rangle_0$; $f= \langle \eta \rangle_0$, $\langle \rho \rangle_0$; $f=\langle \chi \rangle_0$ and $f \gg \langle \chi \rangle_0$ and obtain the spectrum of scalars for each case. In the first and second cases the spectrum of scalars presents a set of new scalars belonging to the electroweak scale, while in the third case all new scalars belong to the 3-3-1 scale and the fourth case all the new scalars have masses lying at $f$ scale. All cases  have a neutral CP-even scalar mimicking the standard Higgs.
\end{abstract}

\maketitle
\section{introduction}
The detection of the standard Higgs boson by ATLAS\cite{ATLAS:2012yve}  and CMS\cite{CMS:2012qbp} and the search for additional Higgs bosons by the LHC, motivates the study of the scalar sector of gauge extensions of the standard model (SM) with the purpose of checking if the scalar spectra of these models contain a CP even with $125$ GeV of mass that mimics the Higgs of the SM and  verify if the spectra of new scalars is reachable by current and future colliders.

The $SU(3)_C \times SU(3)_L \times U(1)_N$  model  with right-handed neutrinos (331RHNs)\cite{Singer:1980sw,Frampton:1992wt,Pisano:1992bxx,Foot:1994ym} is a well-motivated gauge extension of the standard model. It provides an understanding for family replication\cite{Ng:1992st,Liu:1993gy,Pisano:1996ht}, solve the strong CP-problem\cite{Pal:1994ba,Dias:2003iq,Montero:2011tg}, gives an explanation for the pattern of quantization of electric charges\cite{deSousaPires:1998jc,deSousaPires:1999ca} and has a rich phenomenology in what concern flavor physics\cite{Ng:1992st,Liu:1993gy}\cite{GomezDumm:1993oxo,Promberger:2007py, Cabarcas:2009vb,Buras:2012dp,Queiroz:2016gif,Cogollo:2012ek,Machado:2013jca,Benavides:2009cn,Oliveira:2022vjo,CarcamoHernandez:2022fvl}, dark matter \cite{deSPires:2007wat,Mizukoshi:2010ky,Cogollo:2014jia,Montero:2017yvy,Profumo:2013sca,Oliveira:2021gcw} and provide a robust material for collider physics as, for example, new neutral and charged gauge bosons belonging to the TeV scale\cite{Long:1996rfd}. 

 Here we are interested in the Higgs sector of the 331RHNs. Besides this model has received considerable attention in the last decades, its scalar sector is poorly developed and its mass spectra is unknown\cite{Long:1997vbr,Tully:1998wa,Ponce:2002sg,Palcu:2013sfa,Diaz:2003dk} . In this work we develop the scalar sector of the 331RHNs and obtain its spectra of scalars varying the energy scale $f$.

This work is organized as follow: In Sec. II we make a brief review of the scalar sector where we present the texture of all the four mass matrices involving the scalars of the model. In Sec. III we obtain the spectrum of scalars for four scenario characterized by the regime of energy that $f$ is supposed to belong. In Sec. IV we present our conclusions.

\section{General aspect of the scalar sector}
The scalar sector of the 331RHNs is composed by the following three triplets  \cite{Foot:1994ym},
\begin{eqnarray}
\eta = \left (
\begin{array}{c}
\eta^0 \\
\eta^- \\
\eta^{\prime 0}
\end{array}
\right ),\,\rho = \left (
\begin{array}{c}
\rho^+ \\
\rho^0 \\
\rho^{\prime +}
\end{array}
\right ),\,
\chi = \left (
\begin{array}{c}
\chi^0 \\
\chi^{-} \\
\chi^{\prime 0}
\end{array}
\right ),
\label{scalarcont} 
\end{eqnarray}
with $\eta$ and $\chi$ transforming as $(1\,,\,3\,,\,-1/3)$
and $\rho$ as $(1\,,\,3\,,\,2/3)$.  After spontaneous breaking of the symmetry, such content of scalar generates masses for all massive particles of the model including fermions and gauge bosons\footnote{For the development of the other sectors of the model, see Refs.\cite{Long:1995ctv,Long:1997hwz,Cao:2016uur,Oliveira:2022vjo}}.

The most general potential that conserve lepton number involves the following terms\cite{Pal:1994ba}:
\begin{eqnarray} 
V(\eta,\rho,\chi)&=&\mu_\chi^2 \chi^2 +\mu_\eta^2\eta^2
+\mu_\rho^2\rho^2+\lambda_1\chi^4 +\lambda_2\eta^4
+\lambda_3\rho^4+ \nonumber \\
&&\lambda_4(\chi^{\dagger}\chi)(\eta^{\dagger}\eta)
+\lambda_5(\chi^{\dagger}\chi)(\rho^{\dagger}\rho)+\lambda_6
(\eta^{\dagger}\eta)(\rho^{\dagger}\rho)+ \nonumber \\
&&\lambda_7(\chi^{\dagger}\eta)(\eta^{\dagger}\chi)
+\lambda_8(\chi^{\dagger}\rho)(\rho^{\dagger}\chi)+\lambda_9
(\eta^{\dagger}\rho)(\rho^{\dagger}\eta) \nonumber \\
&&-\frac{f}{\sqrt{2}}\epsilon^{ijk}\eta_i \rho_j \chi_k +\mbox{H.c.}\,.
\label{potentialII}
\end{eqnarray}

In order to avoid spontaneous breaking of the lepton number, we assume that only  $\eta^0\,\,, \rho^0$, and $\chi^{\prime 0}$ develop VEV\footnote{For the case where the other neutral scalars develop VEVs, see Ref. \cite{deSPires:2003wwk}}. Shifting the fields  in the usual way,
\begin{equation}
  \eta^0\,\,, \rho^0\,\,, \chi^{\prime 0} =\frac{1}{\sqrt{2}}(v_{\eta\,, \rho\,,\chi^{\prime}}+R_{_{\eta\,, \rho\,,\chi^{\prime}}}+iI_{_{\eta\,, \rho\,,\chi^{\prime}}})\,,
\end{equation}
the potential above provides the following set of  equations that guarantee the potential develop a  minimum
\begin{eqnarray}
 &&\mu^2_\chi +\lambda_1 v^2_{\chi^{\prime}} +
\frac{\lambda_4}{2}v^2_\eta  +
\frac{\lambda_5}{2}v^2_\rho-\frac{f}{2}\frac{v_\eta v_\rho}
{ v_{\chi^{\prime}}}=0,\nonumber \\
&&\mu^2_\eta +\lambda_2 v^2_\eta +
\frac{\lambda_4}{2} v^2_{\chi^{\prime}}
 +\frac{\lambda_6}{2}v^2_\rho -\frac{f}{2}\frac{v_{\chi^{\prime}} v_\rho}
{ v_\eta} =0,
\nonumber \\
&&
\mu^2_\rho +\lambda_3 v^2_\rho + \frac{\lambda_5}{2}
v^2_{\chi^{\prime}} +\frac{\lambda_6}{2}
v^2_\eta-\frac{f}{2}\frac{v_\eta v_{\chi^{\prime}}}{v_\rho} =0.
\label{mincond} 
\end{eqnarray}
After solving the minimum conditions of the potential, we are able to  obtain the mass matrices of the scalars of the model. We remember that  $v^2_\eta + v^2_\rho=(246)^2$GeV$^2$, while $v_{\chi^{\prime}}$ characterize the energy scale of the breaking of the $SU(3)_L \times U(1)_N$ symmetry.  Current LHC bounds imposes  $v_{\chi^{\prime}} \geq 4$TeV\cite{Coutinho:2013lta,Alves:2022hcp}.

The potential above and the constraints in Eq. (\ref{mincond}) provide the following  mass matrix for the CP-even neutral scalars  according the basis $(R_{\chi^{\prime}}, R_\eta,R_\rho)$:
\begin{equation}
M_R^2=
\begin{pmatrix}
\lambda_1 v^2_{\chi^{\prime}}+fv_\eta v_\rho/4v_{\chi^{\prime}} & \lambda_4 v_{\chi^{\prime}}  v_\eta/2- f v_\rho/4 &  \lambda_5 v_{\chi^{\prime}}  v_\rho/2- f v_\eta/4\\
\lambda_4 v_{\chi^{\prime}}  v_\eta/2- f v_\rho/4 & \lambda_2 v^2_\eta+ fv_{\chi^{\prime}} v_\rho/4v_\eta & \lambda_6 v_\eta v_\rho/2-f v_{\chi^{\prime}}/4\\
\lambda_5 v_{\chi^{\prime}}  v_\rho/2- f v_\eta/4 &  \lambda_6 v_\eta v_\rho/2-f v_{\chi^{\prime}}/4 &  \lambda_3 v^2_\rho+ fv_{\chi^{\prime}} v_\eta/4v_\rho
\end{pmatrix},
\label{masseven}
\end{equation}
After diagonalization, the eigenvalues of this matrix
cannot assume a simple analytic form due to the parameter $f$. Hence, we can only find eigenvalues and eigenvectors associated to $M_R^2$ after assuming some approximations among $f$ and the others energy scales $v_{\chi^{\prime}}$, $v_\eta$ and $v_\rho$.  Here we consider  four scenarios  we think that are important.

Now, let us move to the CP-odd scalars. Considering the basis $(I_{\chi^{\prime}}, I_\eta, I_\rho)$, we obtain the following mass matrix for these scalars:
\begin{equation}
M_I^2=
\begin{pmatrix}
fv_\eta v_\rho/4v_{\chi^{\prime}} &  f v_\rho/4 &  f v_\eta/4\\
f v_\rho/4 &  fv_{\chi^{\prime}} v_\rho/4v_\eta & f v_{\chi^{\prime}}/4\\
 f v_\eta/4 &  f v_{\chi^{\prime}}/4 &  fv_{\chi^{\prime}} v_\eta/4v_\rho
\end{pmatrix}.
\label{MI}
\end{equation}
This matrix allows an analytical solutions to their eigenvalues and eigenvectors.  Assuming $v_{\chi^{\prime}}\gg v_\eta\,,\,v_\rho$ we get two null eigenvalues, which correspond to the eigenstates $I_{\chi^{\prime}}$ and $G= \frac{v_\eta}{\sqrt{v^2_\eta + v^2_\rho}} I_\eta -\frac{v_\rho}{\sqrt{v^2_\eta + v^2_\rho}}I_\rho$, and  a massive one with mass expression given by  $m^2 _A=\frac{f v_{\chi^{\prime}}}{4}(\frac{v_\eta v_\rho}{v^2_{\chi^{\prime}}}+\frac{v_\eta}{v_\rho}+\frac{v_\rho}{v_\eta})$ corresponding to the eigenstate  $A= \frac{v_\rho}{\sqrt{v^2_\eta + v^2_\rho}} I_\eta +\frac{v_\eta}{\sqrt{v^2_\eta + v^2_\rho}}I_\rho$. Observe that if $f=0$ then $m_A=0$ which means that $A$ is a pseudo Goldstone associate to some  $U(1)_X$ global symmetry. In fact it was showed in \cite{Pal:1994ba} that $U(1)_X$ is the Peccei-Quinn symmetry and $A$ is the Weinberg-Wilczek axion\cite{Weinberg:1977ma,Wilczek:1977pj}. Of course that this Axion must be avoided. The triliner term $f \eta \rho \chi$ is the most economical way of breaking explicitly Peccei-Quinn symmetry and still conserve lepton number\footnote{For the implementation of Peccei-Quinn symmetry with a safe axion in the 331RHNs, see \cite{Dias:2003iq}}.

The other neutral scalars are $\chi^0$ an $\eta^{\prime 0}$. They carry two units of lepton number each\cite{deSPires:2007wat}. As we assume lepton number is preserved, they do not mix with $\eta^0$, $\rho$ neither $\chi^{\prime 0}$. Considering the  basis $(\chi^{ 0} , \eta^{\prime 0})$ we obtain the following mass matrix

\begin{equation}
M^2_{\chi \eta^{\prime}}=
\begin{pmatrix}
\lambda_7v^2_\eta/4 +f v_\eta v_\rho/4v_{\chi^{\prime}} & -\lambda_7 v_\eta v_{\chi^{\prime}}/4-fv_\rho/4\\
-\lambda_7 v_\eta v_{\chi^{\prime}}/4-fv_\rho/ &  \lambda_7 v^2_{\chi^{\prime}}/4 +f v_\rho v_{\chi^{\prime}}/4v_\eta
\end{pmatrix},
\label{MI2}
\end{equation}

After diagonalizing it we obtain  $G_3=\frac{v_{\chi^{\prime}}}{\sqrt{v^2_\eta + v^2_{\chi^{\prime}}}}\chi^0 +\frac{v_\eta}{\sqrt{v^2_\eta + v^2_{\chi^{\prime}}}}\eta^{\prime 0} $ that  is the Goldstone eaten by the non-hermitian gauge bosons $U ^0$ and $U^{0 \dagger}$. The other neutral scalar is $G_4=-\frac{v_\eta}{\sqrt{v^2_\eta + v^2_{\chi^{\prime}}}}\chi^0 +\frac{v_{\chi^{\prime}}}{\sqrt{v^2_\eta + v^2_{\chi^{\prime}}}}\eta^{\prime 0}$ whose mass is given by the expression $m^2_{G_4}=\frac{\lambda_7}{4}(v_{\chi^{\prime}}^2+ v^2_\eta)+\frac{fv_\rho}{4}(\frac{v_{\chi^{\prime}}}{v_\eta}+\frac{v_\eta}{v_{\chi^{\prime}}})$. This mass  depends of $f$ and $v_{\chi^{\prime}}$ which means that this scalar is a heavy particle whatever is the approximation assumed  on $f$.

Now, let us consider the charged scalars. According to the basis $(\chi^- , \rho^{\prime -}, \eta^- , \rho ^-)$, we obtain the following mass matrix for them:

\begin{equation}
M_C^2=
\begin{pmatrix}
\lambda_8 v^2_\rho/2+f v_\eta v_\rho/2 v_{\chi^{\prime}} & \lambda_8 v_\rho v_{\chi^{\prime}}/2 +fv_\eta/2 & 0 & 0 \\
\lambda_8 v_\rho v_{\chi^{\prime}}/2 +fv_\eta/2 & \lambda_8 v^2_{\chi^{\prime}}/2 +f v_\eta v_{\chi^{\prime}}/2 v_\rho & 0 & 0\\
0 & 0 & \lambda_9 v^2_\rho /2 +f v_\rho v_{\chi^{\prime}}/2 v_\eta & \lambda_9 v_\rho v_\eta /2 +f  v_{\chi^{\prime}}/2 \\
0 & 0 &  \lambda_9 v_\rho v_\eta /2 +f  v_{\chi^{\prime}}/2 & \lambda_9 v^2_\eta /2 +f v_\eta v_{\chi^{\prime}}/2 v_\rho
\end{pmatrix}.
\label{M+}
\end{equation}

 We remember that $\chi^- , \rho^{\prime -}$ carry two units of lepton number each\cite{deSPires:2007wat}. Thus lepton number conservation explains why  $\chi^- , \rho^{\prime -}$ do not mix with $\eta^- , \rho ^-$. After diagonalizing this matrix we obtain two Goldstone $G_1^+=-\frac{v_\rho}{\sqrt{v^2_\eta + v^2_\rho}}\rho^+ +\frac{v_\eta}{\sqrt{v^2_\eta + v^2_\rho}} \eta^+$ and $ G_2^{ +}= \frac{v_{\chi^{\prime}}}{\sqrt{v^2_\eta + v^2_{\chi^{\prime}}}}\chi^+\frac{v_{\eta}}{\sqrt{v^2_\eta + v^2_{\chi^{\prime}}}}\rho^{\prime +}$ that will be eaten by $W^{\pm}$ and $W^{\prime \pm}$. The remain two heavy charged scalars are $h_1^+=\frac{v_\eta}{\sqrt{v^2_\eta + v^2_\rho}}\rho^+ +\frac{v_\rho}{\sqrt{v^2_\eta + v^2_\rho}} \eta^+ $ with mass $m_{h_1^{\pm}}^2=\frac{1}{2}(f v_{\chi^{\prime}}+\lambda_9 v_\eta v_\rho)(\frac{v_\eta}{v_\rho}+\frac{v_\rho}{v_\eta}) $  and $h_2^+=-\frac{v_{\eta}}{\sqrt{v^2_\eta + v^2_{\chi^{\prime}}}}\chi^++ \frac{v_{\chi^{\prime}}}{\sqrt{v^2_\eta + v^2_{\chi^{\prime}}}}\rho^{\prime +} $   with mass  $m^2_{h_2^+}=\frac{1}{2}(f v_{\eta}+\lambda_8 v_\rho v_{\chi^{\prime}})(\frac{v_{\chi^{\prime}}}{v_\rho} +\frac{v_\rho}{v_{\chi^{\prime}}}) $. Observe that the dependence of $m^2_{h_2^{\pm}}$ on $f$ and $v_{\chi^{\prime}}$ is such that $h^{\pm}_2$ is a heavy particle whatever is the approximation on $f$, too.

Summarizing, the spectrum of scalars of the model possess  four heavy scalars, namely $H$, $G_4$ and $h^{\pm}_2$ and other four scalars $h_2$, $A$ and $h^{\pm}_1$  have their masses determined by $f$. So in order to know to which energy scale these particles belongs to, we first have to determine $f$ . As there are no constraints on $f$,  we will discuss four energy regimes for $f$.

\section{The Higgs spectra}

\subsection{Spectrum I}
Here we consider the case in which $f$ is supposed to belong to low energy scales, which is achieved with the assumption $v_{\chi^{\prime}} \gg v_\eta = v_\rho =v\gg f$. In order to see the impact of this hierarchy on the mass matrix $M_R^2$, let us write it in the following way:
\begin{equation}
M_R^2 =v^2_{\chi^{\prime}}
\begin{pmatrix}
\lambda_1 +\frac{f}{4v_{\chi^{\prime}}}\delta^2  &\frac{ \lambda_4}{2} \delta- \frac{f}{4v}\delta^2&  \frac{\lambda_5}{2} \delta- \frac{f}{v}\delta^2\\
 \frac{ \lambda_4}{2} \delta- \frac{f}{4v}\delta^2&  \frac{f}{4v_{\chi^{\prime}}}+\lambda_2 \delta^2 &  \frac{\lambda_6}{2}\delta^2-\frac{f}{4v_{\chi^{\prime}}}\\
 \frac{ \lambda_5}{2} \delta- \frac{f}{4v}\delta^2&  \frac{\lambda_6}{2}\delta^2-\frac{f}{4v_{\chi^{\prime}}} &  \frac{f}{4v_{\chi^{\prime}}}+\lambda_3 \delta^2
\end{pmatrix},
\label{P1}
\end{equation}
where $\delta =\frac{v}{v_{\chi^{\prime}}}$. For $\delta \ll 1$, the only relevant mixings appear between $R_{\chi^{\prime}}-R_\eta$ and  $R_{\chi^{\prime}}-R_\rho$ which means that only the following terms matter in the matrix $M_R^2$:
\begin{equation}
M_R^2 \approx
v^2_{\chi^{\prime}}\begin{pmatrix}
\lambda_1  &\frac{ \lambda_4}{2} \delta&  \frac{\lambda_5}{2} \delta\\
 \frac{ \lambda_4}{2} \delta &  0 &  0\\
 \frac{ \lambda_5}{2} \delta & 0 & 0
\end{pmatrix}
\end{equation}

The eigenvectors associated to this matrix are:
\begin{eqnarray}
&&H\cong R_\chi,\nonumber \\
  && h_1\cong\frac{\lambda_4}{\sqrt{\lambda_4^2+\lambda_5^2}}R_\eta +\frac{\lambda_5}{\sqrt{\lambda_4^2+\lambda_5^2}}R_\rho ,\nonumber \\
  &&
  h_2\cong-\frac{\lambda_5}{\sqrt{\lambda_4^2+\lambda_5^2}}R_\eta +\frac{\lambda_4}{\sqrt{\lambda_4^2+\lambda_5^2}}R_\rho .
  \label{S1}
\end{eqnarray}
For sake of simplicity we  take $\lambda_4=\lambda_5$. In order to obtain the eigenvalues associated to these eigenvectors we solve individually the  equations
\begin{equation}
    (M_R^2-m^2_H I)H=0,\,\,\,\,\,\,\,(M_R^2-m^2_{h_1} I)h_1=0,\,\,\,\,\,\,(M_R^2-m^2_{h_2} I)h_2=0,
\end{equation}
for $M_R^2$ given in Eq. (\ref{P1}) with $I$ being the identity matrix. For this we obtain,
\begin{equation}
    m^2_H= \lambda_1 v^2_{\chi^{\prime}}\,\,\,,\,\,\,m^2_{h_1}=(\lambda_2+\frac{\lambda_6}{2})v^2\,\,\,,\,\,\, m^2_{h_2}=(\lambda_2-\frac{\lambda_6}{2})v^2+fv_{\chi^{\prime}}
    \label{S2}
\end{equation}
We recognize $h_1$ as the scalar that will play the role of the Higgs. $H$ is a heavy Higgs associated to the 3-3-1 energy scale and $h_2$ is a second Higgs with mass around the electroweak scale for tiny $f$.

In regard to the pseudoscalar, $A$, we have that its mass expression displayed in the previous section take the following form $m^2_A \approx \frac{fv_{\chi^{\prime}}}{2}$
 with $A= \frac{1}{\sqrt{2}}I_\eta +\frac{1}{\sqrt{2}}I_\rho$.

One concrete way to investigate the energy scale of $f$ is looking for constraints on $A$. Firstly,  as the decay channel $h_1 \rightarrow AA$ has not being detected, this demands  $m_A > 62.5$GeV which implies, for $v_{\chi^{\prime}}=10^4$GeV,  in $f>0.8 $GeV.   According to Eq. (\ref{S1}) and $v=174$ GeV, a Higgs with mass of $125$ GeV demands $\lambda_2+\frac{\lambda_6}{2}\approx 0.51$. In view of this it is reasonable to assume $\lambda_2-\frac{\lambda_6}{2}<0.51$. Let us be conservative and take $\lambda_2-\frac{\lambda_6}{2}=0.5$. We, then, have $m_{h_2}\approx 105$ GeV. Thus, according to this illustrative case  the 331RHNs may have a second Higgs lighter than the standard one. This is a very appealing scenario for physics colliders

 For the neutral scalar $G_4$, its mass take the expression  $m^2_{G_4}\approx \frac{\lambda_7}{4}v^2_{\chi^{\prime}}+\frac{fv_{\chi^{\prime}}}{4}$ with $G_4 \approx \eta^{\prime 0}$. This scalar is heavy because of the appearance of $v_{\chi^{\prime}}$ in its mass expression.
 
 For the charged scalars, we have  $h_2^+\approx \rho^{\prime +} $   with mass  $m^2_{h_2^+}=\frac{1}{2}(f v_{\chi^{\prime}}+\lambda_8 v^2_{\chi^{\prime}}) $ that is a heavy scalar with mass belonging to the 3-3-1 scale. The other charged scalar   $h_1^+=\frac{1}{\sqrt{2}}\rho^+ +\frac{1}{\sqrt{2}} \eta^+ $ has  mass $m_{h_1^+}^2=(f v_{\chi^{\prime}}+\lambda_9 v^2)$.   For $f=0.8$ and $\lambda_9=1$ the model predicts $m_{h^{\pm}_1}\approx 196$ GeV. Then the model  may  have a charged scalar with mass belonging to the electroweak scale that may give interesting contribution to flavor physics as, for example,  $B$ meson  decays \cite{Cherchiglia:2022zfy,Okada:2016whh,Buras:2013dea,Promberger:2008xg,Rodriguez:2004mw}.
 
 In summary, the 331RHNs with tiny $f$ has a Higgs with mass of $125$ GeV and  allows a spectrum of scalars involving four heavy scalars namely,  $H$, $G_4$ and $h_2^{\pm}$ and four  scalars $h_2$,  $h_1^{\pm}$ and $A$ with mass belonging to the electroweak scale with $h_2$ could being lighter than the standard Higgs. This case recovers the two Higgs doublet model and has a strong phenomenological appeal.

 \subsection{Spectrum II}
 We refer to spectrum II the case characterized by the assumption $v_{\chi^{\prime}} \gg  v_\eta = v_\rho = f=v$. According to this approximation, the dominant terms in $M_R^2$ are
\begin{equation}
M_R^2 =v^2_{\chi^{\prime}}
\begin{pmatrix}
\lambda_1 +\frac{1}{4}\delta^3  &\frac{ \lambda_4}{2} \delta- \frac{1}{4}\delta^2&  \frac{\lambda_5}{2} \delta- \frac{1}{4}\delta^2\\
 \frac{ \lambda_4}{2} \delta- \frac{1}{4}\delta^2&  \frac{1}{4}\delta+\lambda_2 \delta^2 &  \frac{\lambda_6}{2}\delta^2-\frac{1}{4}\delta\\
 \frac{\lambda_5}{2} \delta- \frac{1}{4}\delta^2&   \frac{\lambda_6}{2}\delta^2-\frac{1}{4}\delta &  \frac{1}{4}\delta+\lambda_3 \delta^2
\end{pmatrix}\approx v^2_{\chi^{\prime}}
\begin{pmatrix}
\lambda_1   &\frac{ \lambda_4}{2} \delta &  \frac{\lambda_5}{2} \delta\\
 \frac{ \lambda_4}{2} \delta&  \frac{1}{4}\delta &  -\frac{1}{4}\delta\\
 \frac{\lambda_5}{2} \delta&  -\frac{1}{4}\delta &  \frac{1}{4}\delta
\end{pmatrix},
\label{MuII}
\end{equation}

Following the same procedure of the previous case, we obtain:

\begin{eqnarray}
   && m^2_{h_1}\simeq (\lambda_2+\frac{\lambda_6}{2})v^2\,\,\,\rightarrow \,\,\,h_1=\frac{R_\eta}{\sqrt{2}}+\frac{R_\rho}{\sqrt{2}},\nonumber \\
    &&m^2_{h_2} \simeq (\lambda_2-\frac{\lambda_6}{2})v^2+\frac{vv_{\chi^{\prime}}}{2}\simeq \frac{vv_{\chi^{\prime}}}{2}\,\,\,\rightarrow \,\,\,h_2=-\frac{R_\eta}{\sqrt{2}}+\frac{R_\rho}{\sqrt{2}},\nonumber \\
    && m_H^2 \simeq \lambda_1 v_{\chi'}^2/2 \,\,\, \rightarrow \,\,\, H\simeq R_ {\chi}.
\end{eqnarray}
The other neutral scalar, $G_4$, has mass given by $m^2_{G_4}= \frac{\lambda_7}{4}(v^2_{\chi^{\prime}}+v^2)+\frac{vv_{\chi^{\prime}}}{4}$. The pseudoscalar, $A$, get the following mass expression $m^2_A=\frac{vv_{\chi^{\prime}}}{2}$. The two  charged scalars get the following mass expression $m^2_{h^+_1}=\lambda_9 v^2+vv_{\chi^{\prime}}\simeq vv_{\chi^{\prime}}$ and $m^2_{h^+_2}=\frac{1}{2}(\lambda_8v^2_{\chi^{\prime}}+v^2)\simeq \frac{1}{2}\lambda_8v^2_{\chi^{\prime}} $.

In summary, in this case the Higgs  $h_2$, $A$ and  $h_1^{\pm}$ still may develop values around the electroweak scale while $H\,\,,\,\, h_{2}^{\pm}\,\,\,,\,\, G_4$   are heavy particles with mass belonging to the 3-3-1 scale. As illustrative example, fixing  $\lambda_8=\lambda_9=1$ we get $m_{h_2}=m_A \approx 660$ GeV, $m_{h^+_1}=1319$ GeV and  $m_{h^+_2}=5000$ GeV.  This is also an interesting scenario from the point of view of phenomenology and can be probed at the LHC.

 \subsection{Scenario III}
 We refer to spectrum III the case in which  $v_{\chi^{\prime}}=f \gg  v_\eta = v_\rho =v$. This case has being previously studied in the literature\cite{deSPires:2007wat} and the dominant terms in $M_R^2$ are
 \begin{equation}
M_R^2 = v^2_{\chi^{\prime}}
\begin{pmatrix}
\lambda_1 +\frac{1}{4}\delta^2  &(\frac{ \lambda_4}{2} - \frac{1}{4})\delta&  (\frac{\lambda_5}{2} - \frac{1}{4})\delta\\
 (\frac{ \lambda_4}{2} - \frac{1}{4})\delta&  \frac{1}{4}\delta+\lambda_2 \delta^2 &  \frac{\lambda_6}{2}\delta^2-\frac{1}{4}\\
 (\frac{\lambda_5}{2} - \frac{1}{4})\delta&   \frac{\lambda_6}{2}\delta^2-\frac{1}{4} &  \frac{1}{4}+\lambda_3 \delta^2
\end{pmatrix}
\approx v^2_{\chi^{\prime}} \begin{pmatrix}
\lambda_1 +\frac{1}{4}\delta^2  &0&  0\\
 0 &  \frac{1}{4}\delta+\lambda_2 \delta^2 &  -\frac{1}{4}\\
0 & -\frac{1}{4} &  \frac{1}{4}+\lambda_3 \delta^2
\end{pmatrix}
\label{MuIII}
\end{equation}
Following the procedure above we obtain
 \begin{eqnarray}
    && m^2_H=\lambda_1 v^2_{\chi^{\prime}} \,\,\, \rightarrow H \approx R_ \chi,\nonumber \\
    &&m^2_{h_2} \approx \frac{1}{2}v^2_{\chi^{\prime}}+(\lambda_2 + \lambda_ 3-\lambda_6)v^2, \,\,\,\rightarrow h_2 \approx -\frac{1}{\sqrt{2}}R_\eta +\frac{1}{\sqrt{2}}R_\rho,\nonumber \\
    &&m^2_{h_1} \approx (\lambda_2 + \lambda_ 3+\lambda_6)v^2, \,\,\,\rightarrow h_1 \approx \frac{1}{\sqrt{2}}R_\eta +\frac{1}{\sqrt{2}}R_\rho.
 \end{eqnarray}

The neutral scalar $G_4$ get the mass  $m^2_{G_4}=\frac{\lambda_7}{4}(v^2_{\chi^{\prime}}+v^2)+\frac{v^2_{\chi^{\prime}}}{4}\approx \frac{v^2_{\chi^{\prime}}}{4} $. The same for the pseudoscalar $m^2_A=\frac{v^2_{\chi^{\prime}}}{4}$. The two charged scalars get also heavy with $m^2_{h^{\pm}_2}=\frac{1}{2}(\lambda_8 v^2_{\chi^{\prime}}+v^2_{\chi^{\prime}})$ and $m^2_{h^{\pm}_1}=(\lambda_9v^2+v^2_{\chi^{\prime}})$. In this case, all scalars, except the Higgs, get mass at the 3-3-1 scale. This scenario is not so phenomenologically interesting for present collider physics as the previous two.

 \subsection{Spectrum IV}
 We now discuss the case characterized by the approximation $f> v_{\chi^{\prime}} \gg  v_\eta , v_\rho $.  For $f$ belonging to such regime of energy one has $m^2_A=\frac{fv_{\chi^{\prime}}}{2}$, $m^2_{G_4}\approx \frac{\lambda_7}{4}f v_{\chi^{\prime}}$, while $m^2_{h^{\pm}_2}=\frac{1}{2}f v_{\chi^{\prime}}$ and $m^2_{h^{\pm}_1}=f v_{\chi^{\prime}}$. So, all these scalars  are heavy particles, since they are proportional to the product $f v_{\chi^{\prime}}$ and can not be probed at present collider.

  Now let us discuss  the eigenvalues of $M_R^2$. For  $f$ belonging to high energy scale the consequence is the fact that  it dominates the elements of the matrix $M_R^2$. Our worry here  was if in such regime of energy for $f$ the model  contains yet a Higgs-like particle with mass of $125$ GeV. 
  
\begin{figure}
    \begin{subfigure}{.8\textwidth}
        \centering   \includegraphics[width=.8\linewidth]{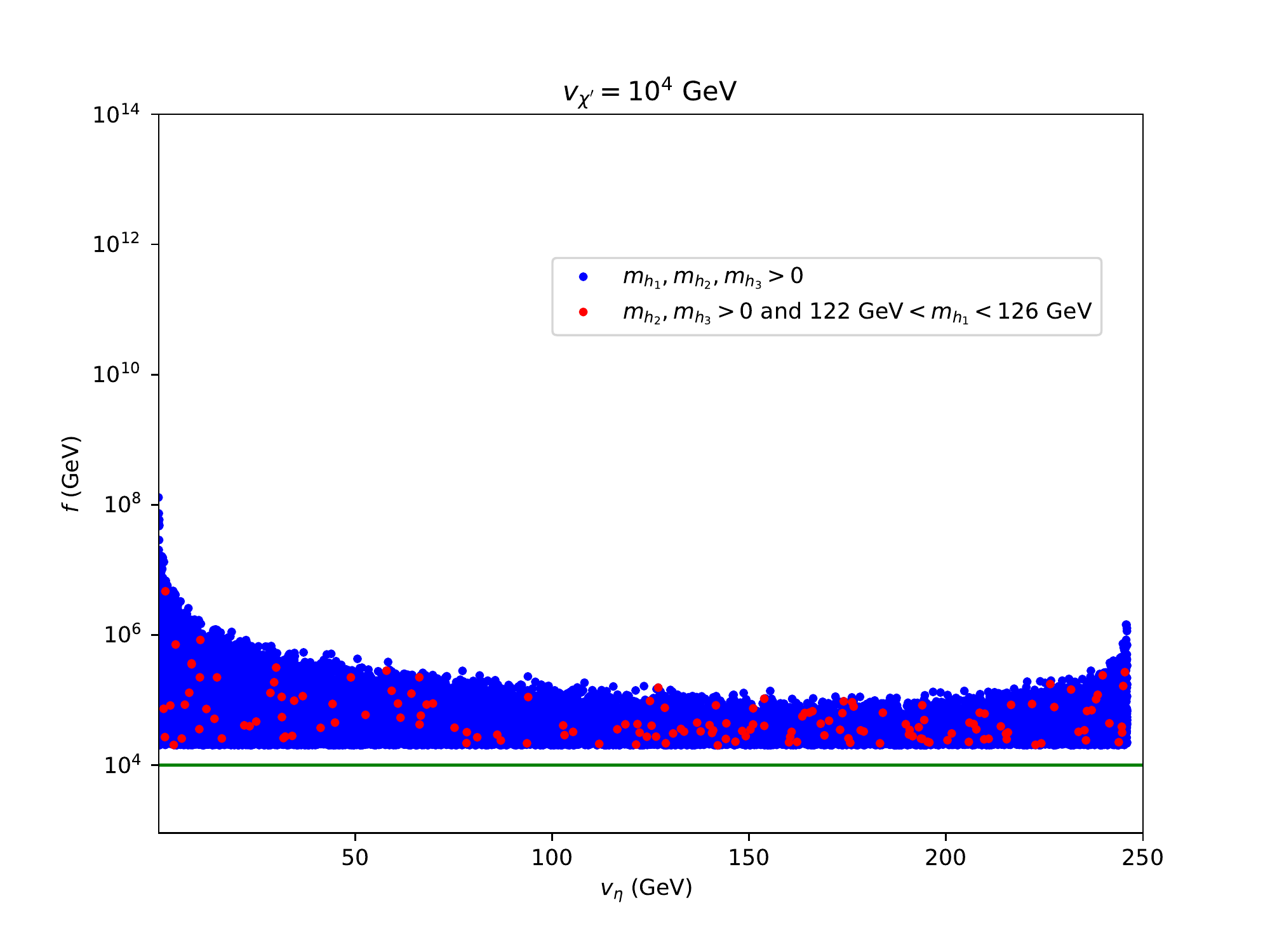}
      \caption{}
      \label{fveta1}
    \end{subfigure}%
    \par\bigskip 
    \begin{subfigure}{.8\textwidth}
       \centering     \includegraphics[width=.8\linewidth]{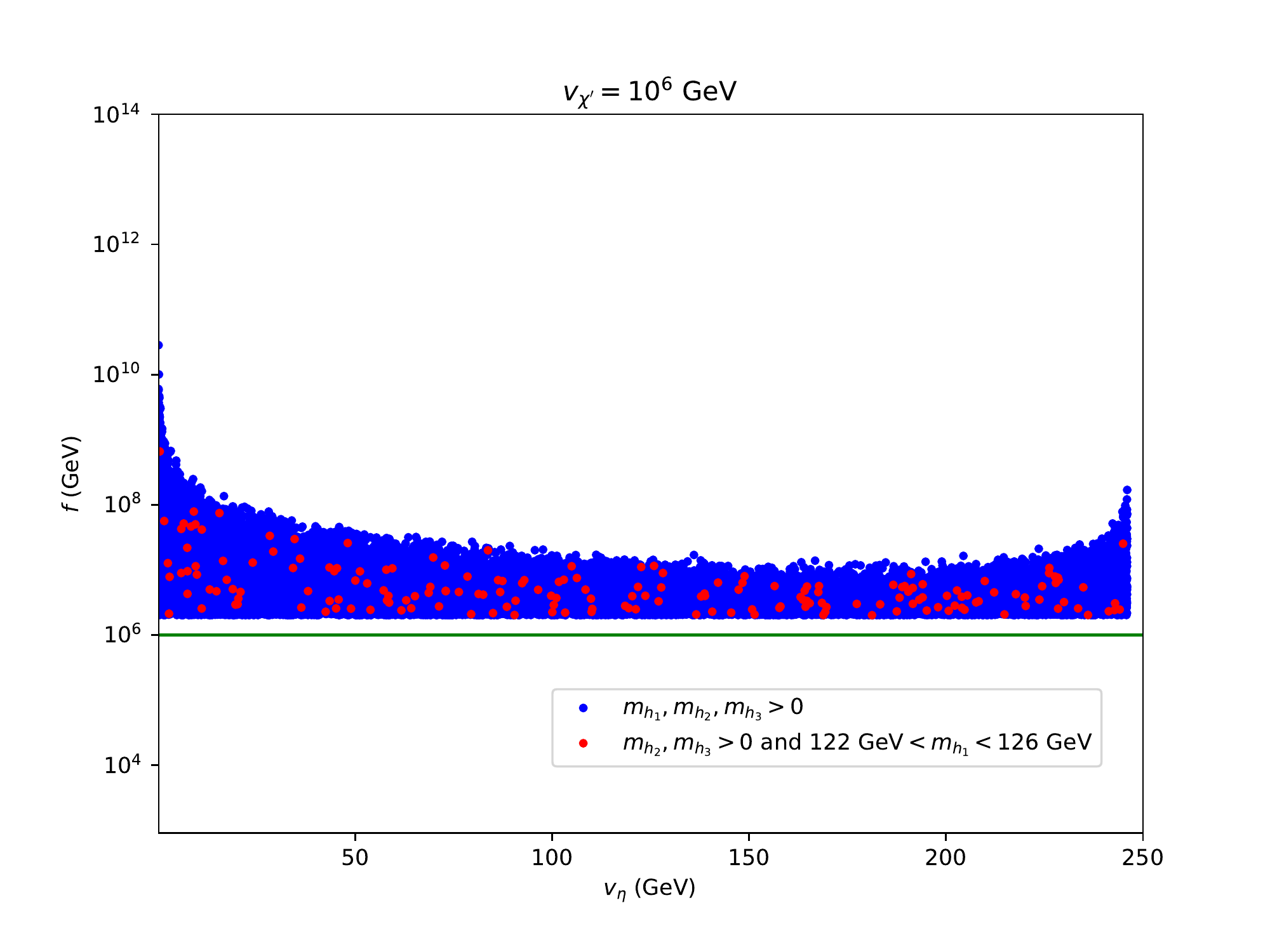}
      \caption{}
      \label{fveta2}
    \end{subfigure}
\caption{ Here, for a fixed $v_{\chi^\prime}$, we are varying all other parameters of the potential. All quartic couplings $\lambda_{i}$'s are varying randomly between $-\sqrt{4 \pi }$ and $\sqrt{4 \pi }$, $f$ is varying between $2\times v_{\chi^\prime}$ and $10^{14}$ GeV, $v_\eta$ is varying between $0.01$ GeV and $245.99$ GeV, and $v_\rho$ respects the relation $v_\rho = \sqrt{246^2\;\mathrm{ GeV}^2 - v_\eta^2}$. The blue dots represents the allowed values for the two parameters such that the three eigenvalues of the mass matrix are positive, and the red dots represents the values of $f$ and $v_\eta$ such that $122\;\mathrm{GeV}<m_{h_1}<126\;\mathrm{GeV}$. The green horizontal line means the fixed value of $v_{\chi^\prime}$, (a) $v_{\chi^\prime}=10^4$ GeV and (b) $v_{\chi^\prime}=10^6$ GeV.}
\label{fveta}
\end{figure}

\begin{figure}
       \centering    \includegraphics[scale=0.7]{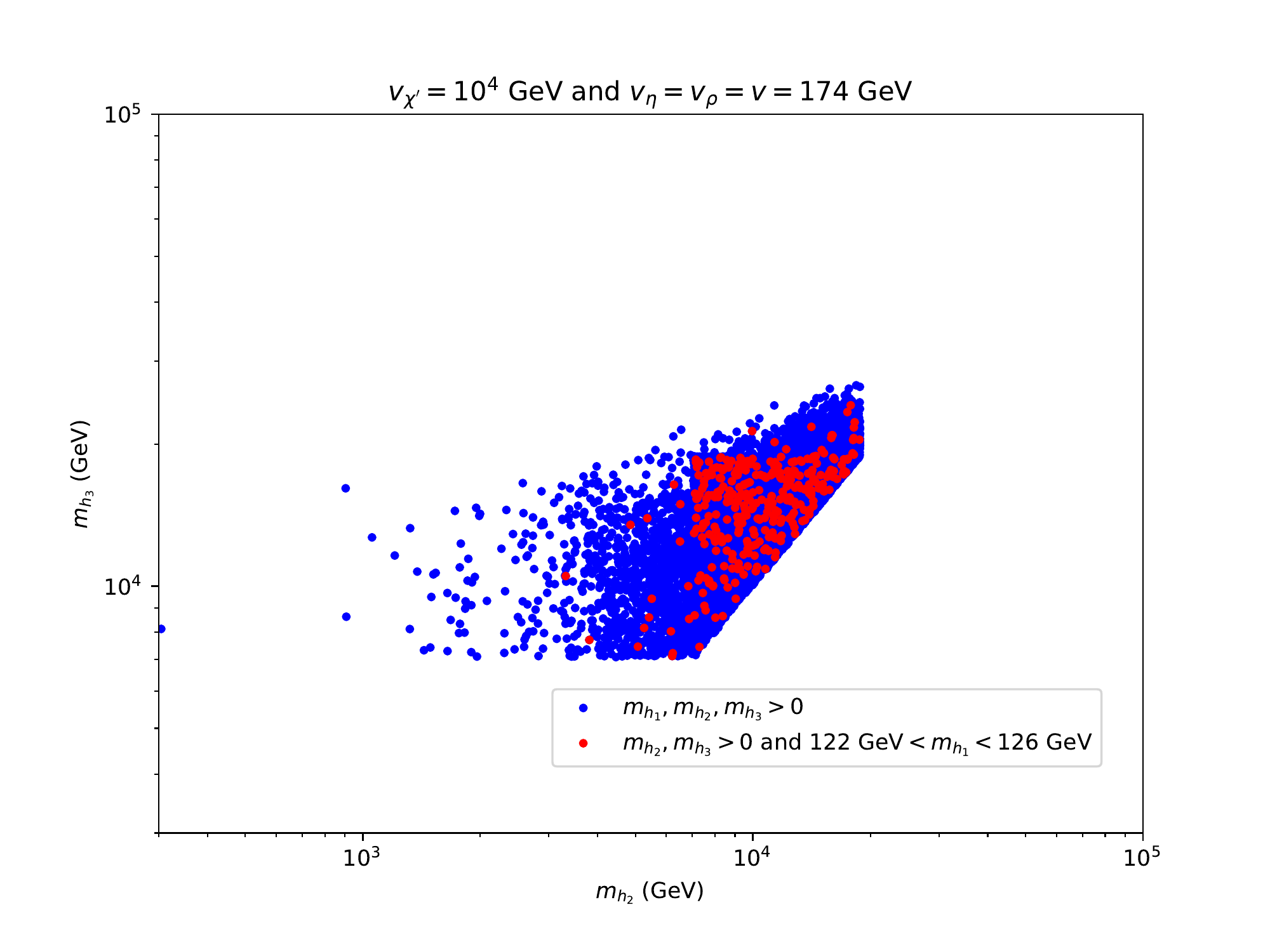}
      \caption{ Here, for a fixed $v_{\chi^\prime}=10^4$ GeV and $v_\eta=v_\rho = v=174$ GeV, we are varying all other parameters of the potential. All quartic couplings $\lambda_{i}$'s are varying randomly between $-\sqrt{4 \pi }$ and $\sqrt{4 \pi }$, $f$ is varying between $2\times v_{\chi^\prime}$ and $10^{14}$ GeV. Hence, we are observing the interdependence of the CP even scalars masses  $m_{h_2}$ and $m_{h_3}$, such that $m_{h_1}$ assumes values that are proportional with the Higgs mass.}
      \label{mh1}
    \end{figure}

\begin{figure}
    \begin{subfigure}{.7\textwidth}
      \centering
      \includegraphics[width=.8\linewidth]{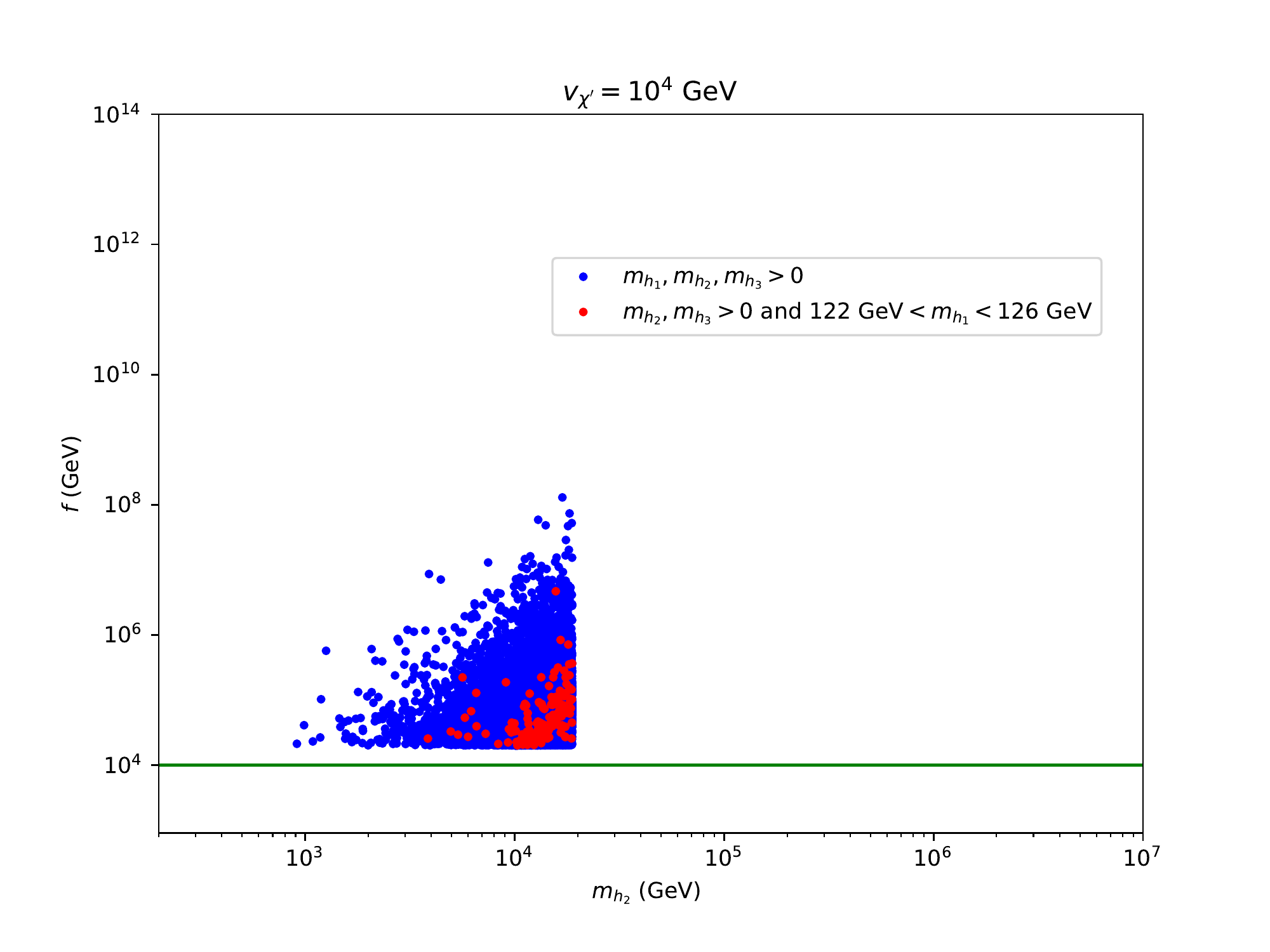}
      \caption{}
      \label{fmh21}
    \end{subfigure}%
    \par\bigskip 
    \begin{subfigure}{.7\textwidth}
      \centering
      \includegraphics[width=.8\linewidth]{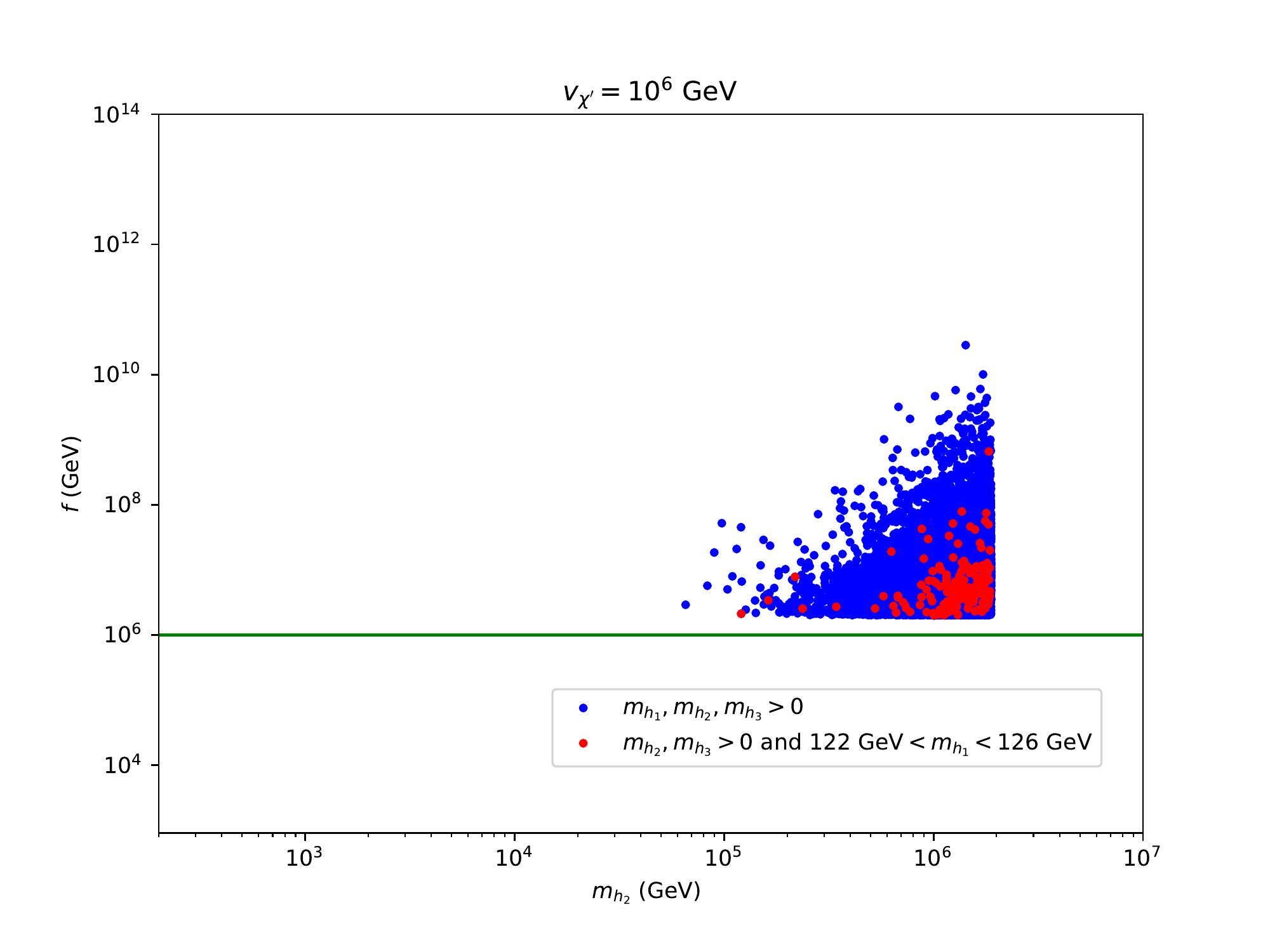}
      \caption{}
      \label{fmh22}
    \end{subfigure}
\caption{As before, for a fixed $v_{\chi^\prime}$,  we are varying all other parameters of the potential.  In order to compare the mass of the intermediate mass Higgs $m_{h_2}$ with $f$. All quartic couplings $\lambda_{i}$'s are varying randomly between $-\sqrt{4 \pi }$ and $\sqrt{4 \pi }$, $f$ is varying between $2\times v_{\chi^\prime}$ and $10^{14}$ GeV, $v_\eta$ is varying between $0.01$ GeV and $245.99$ GeV, and $v_\rho$ respects the relation $v_\rho = \sqrt{246^2\;\mathrm{ GeV}^2 - v_\eta^2}$ . The blue dots represents the allowed values for the two parameters such that the three Higgs masses are positive, and the red dots represents the values of $f$ and $v_\eta$ such that $122\;\mathrm{GeV}<m_{h_1}<126\;\mathrm{GeV}$. Here there is a pattern  $\sqrt{v_{\chi^\prime} f} \sim m_{h_2}$. The green horizontal line means the fixed value of $v_{\chi^\prime}$, (a) $v_{\chi^\prime}=10^4$ GeV and (b) $v_{\chi^\prime}=10^6$ GeV.}
\label{fmh2}
\end{figure}

  We stress here that, in this particular case, the eigenvalues are very sensitive to any  approximation done in the elements of $M^2_R$. We will , then, consider the full  mass matrix without make any approximation on their elements
  \begin{equation}
M_R^2=
\begin{pmatrix}
\lambda_1 v^2_{\chi^{\prime}}+fv_\eta v_\rho/4v_{\chi^{\prime}} & \lambda_4 v_{\chi^{\prime}}  v_\eta/2- f v_\rho/4 &  \lambda_5 v_{\chi^{\prime}}  v_\rho/2- f v_\eta/4\\
\lambda_4 v_{\chi^{\prime}}  v_\eta/2- f v_\rho/4 & \lambda_2 v^2_\eta+ fv_{\chi^{\prime}} v_\rho/4v_\eta & \lambda_6 v_\eta v_\rho/2-f v_{\chi^{\prime}}/4\\
\lambda_5 v_{\chi^{\prime}}  v_\rho/2- f v_\eta/4 &  \lambda_6 v_\eta v_\rho/2-f v_{\chi^{\prime}}/4 &  \lambda_3 v^2_\rho+ fv_{\chi^{\prime}} v_\eta/4v_\rho
\end{pmatrix}.
\label{masseven4}
\end{equation}
Of course, it is not advantageous to address this case analytically. Therefore, using numerical tools, it is possible to understand qualitatively the relation between each parameters of the potential under such hierarchy among the energy scales that characterize this case. Then, we diagonalize numerically $M_R^2$ using a code written in Mathematica 13  language \cite{Mathematica} with 1 million points for each fixed value of $v_{\chi^\prime}$ and extract from it its eigenvalues and check it if at least one eigenvalue  recovers the mass of the Higgs. Our results are shown in FIGs. \ref{fveta}, \ref{mh1} and \ref{fmh2}.  In those figures we are varying all quartic couplings freely between $-\sqrt{4\pi}$ and $\sqrt{4\pi}$, fixing the vev $v_\chi$ for two different values, $10^4$ GeV and $10^6$ GeV represented by the green horizontal lines and varying $f$ between $ 2 \times v_{\chi^\prime}$ and $10^{14}$ GeV. For FIGs. \ref{fveta} and \ref{fmh2} we are varying $v_\eta$ between 0.01 GeV and 245.99 GeV and respecting the bound $v^2_\eta +v^2_\rho=(246)^2$ GeV$^2$ and for FIG \ref{mh1} we are fixing $v_\eta=v_\rho=v=174$ GeV.  The blue dots in each figure means all allowed values for two given parameters such that the three eigenvalues of the mass matrix are positive and the red dots represents the values of the parameters such that 122 GeV$<m_{h_1}<$126 GeV.

FIG \ref{fveta} (a) and (b) show to us that the Higgs mass is kind of independent of the parameter $v_\eta$ (and consequently $v_\rho$), but it depends on the ratio $f/v_{\chi^\prime}$, as expected.  For the other side, FIG \ref{fmh2} (a) and (b) shows the direct dependence of $m_{h_2}$ with $\sqrt{ v_{\chi^\prime} f}$, meaning that the second Higgs is almost degenerated with $m_A$.
In the other side, FIG \ref{mh1}
represents the case in which $v_\eta =v_\rho = v=$ 174 GeV and a fixed $v_{\chi^\prime}=10^4$ GeV. This figure relates  the two heavy states   $m_{h_2}$ and $m_{h_3}$ such that $m_{h_1}$ assumes values that are proportional with the Higgs mass. Here we can see that, even with such simplification $v_\eta=v_\rho$, there are benchmark points that recovers the standard Higgs mass, as expected from our conclusion after analysing FIG \ref{fveta}.

In TABLE \ref{tabnlei} we present a set of three benchmark points to highlight the  solutions presented in FIG. 1 and FIG. II. Of course that this case is not phenomenologically interesting, since the spectrum of 3-3-1  scalars is composed exclusively by  heavy particles. Therefore, the importance of such analysis is to certify that even in this regime of energy  for $f$ the model contains a neutral scalar with mass of $125$ GeV that recovers the standard-like Higgs.  In other words, the model is phenomenologically viable even  for $f$ belonging to high energy scales\footnote{This set of  values for the $\lambda$'s obey  bound from below conditions found in Refs.\cite{Sanchez-Vega:2018qje,Costantini:2020xrn} }.

 \begin{table}
\begin{tabular}{|| c| c| c| c| c| c| c| c|  c|| c| c| c|| }
 \hline
  $v_{\chi^\prime}$(GeV)&$f$(GeV) & $v_{\eta}$(GeV) & $\lambda_1$ & $\lambda_2$ &$\lambda_3$ & $\lambda_4$ & $\lambda_5$ & $\lambda_6$ & $m_{h_3}$(GeV) & $m_{h_2}$(GeV) & $m_{h_1}$(GeV)\\ [0.5ex]
   \hline\hline
 $10^4$ & 3.47$\times 10^4$ &174.09 & 3.51 & -3.14&3.03 &  1.51 & 2.82& 1.23 & 1.87$\times10^{4}$ & 1.32$\times10^{4}$&125.69 \\
  \hline
 $10^5$  & 7.65$\times10^5$ & 164.81 &  2.17&2.57 &  2.89 & -0.51& 1.99 & -0.59 & 1.96$\times10^{5}$&1.47$\times10^{5}$&124.97\\
  \hline
$10^6$ & 2.54$\times 10^6$  &11.01 &   1.54 & 3.18&0.65 & -1.20 & 1.68& 1.56 & 3.76$\times10^{6}$ & 1.24$\times10^{6}$&125.07\\ [1ex] 
 \hline
\end{tabular}
\caption{\label{tabnlei} Three different benchmark points in order to recovery the Higgs mass $m_{h_1}\approx 125$ GeV, for the hierarchy $f>v_{\chi^{\prime}} \gg  v_\eta , v_\rho$.}
\end{table}
\section{Conclusions}
We developed the Higgs sector of the 331RHNs varying  the parameters of the potential with special emphasis on the energy scale $f$.  The potential developed is one that involves only terms that conserve lepton number. In the low energy regime,  $f<v$ , the spectrum of scalars  possesses a set of scalars with mass at electroweak scale with the possibility of  one CP-even scalar  be lighter  than the standard Higgs. In the second case studied, $f=v_\eta$,  the spectrum of new scalars still possesses a set of scalars with mass lying at the electroweak scale. These two cases are phenomenologically interesting and both can be probed at future colliders. The third case is not so appealing for present collider physics but may be relevant for flavor physics. In  the last case studied $f$ belongs to high energy scales ($f\gg v_{\chi^{\prime}}$). In this case, the spectrum of new scalars  develops masses  around $f$ and $v_{\chi^{\prime}}$ escaping the reach of current colliders. Our main worry for such scenario was the possibility of existence of the standard-like Higgs.  We solved numerically the mass matrix for the CP-even neutral scalars and found a set of solutions where a Higgs of $125$ GeV of mass composes the spectrum of scalars, which guarantee that the model is phenomenologically viable, even for a high $f$.

Finally, although the scalar sector of the 331RHNs is composed by three triplets of scalars,  its potential that preserves lepton number is very simple.
Such simplicity can be understood after looking at the spectra of these new scalars. In general, they can be divided into three subsets,  namely one with masses belonging to the 3-3-1 typical scale of energy $(H\,\,,\,\,G_4\,\,,\,\,h_2^{\pm})$, the second with masses varying accordingly to the explicit violation of Peccei-Quinn global symmetry, ($h_2\,\,,\,\, A \,\,,\ h_1^{\pm}$), and the last one at the electroweak symmetry breaking scale ($h_1$).
Since there is a freedom of choice for the Peccei-Quinn symmetry breaking scale, there is a open window to explore the scalars from the second subset. For example, if $f$ belongs to the electroweak or even lower scales, the particles ($h_2\,\,,\,\, A \,\,,\ h_1^{\pm}$) recovers effectively the two Higgs doublet model extension of the standard model. In the other side, for the case in which $f$ belongs to the 3-3-1 scale or higher, the second set of scalars merge into the 3-3-1 scale, generating a bunch of heavy scalars far from the reach of present experiments.  All this flexibility makes the 331RHNs model a very interesting proposal of new physics beyond the standard model.
 
\section*{Acknowledgments}
C.A.S.P  was supported by the CNPq research grants No. 311936/2021-0 and J.P.P. has received funding/support from the European Union’s Horizon 2020 research and innovation program under the Marie Skłodowska-Curie grant agreement No
860881-HIDDeN.
\bibliography{references}{}
\bibliographystyle{ieeetr}
\end{document}